\documentclass{article}
\usepackage{graphicx}
\usepackage{amsmath, amsthm, latexsym}
\usepackage{amssymb}
\begin{document}

\vspace*{3cm} \thispagestyle{empty}

\vspace{5mm}

\noindent \textbf{\Large Causality Violating Geodesics in Bonnor's Rotating Dust Metric}\\

\noindent \textbf{\normalsize Peter Collas}\footnote{Department of Physics and Astronomy, California
State University, Northridge, Northridge, CA 91330-8268. Email: peter.collas@csun.edu.}
\textbf{\normalsize and David Klein}\footnote{Department of Mathematics, California State University,
Northridge, Northridge, CA 91330-8313. Email: david.klein@csun.edu.}\\

\vspace{4mm} \parbox{11cm}{\noindent{\small We exhibit timelike geodesic paths for a metric, 
introduced by Bonnor [11] and considered also by Steadman [13], and show that coordinate time runs
backward along  a portion of these geodesics.}\vspace{5mm}}\\\vspace{6cm}
\pagebreak

\setlength{\textwidth}{27pc}
\setlength{\textheight}{43pc}
\noindent \textbf{{\normalsize 1. INTRODUCTION}}\\

\noindent Causality violating paths are known to exist in many spacetimes satisfying the Einstein 
field equations, especially those solutions associated with rotating matter or rotating
singularities [1-3] (see also Bonnor [4], and references in [5]). However, examples of causality
violating geodesics are sparse in the  scientific literature. Chandrasekhar and Wright [6] showed that
the G\"odel spacetime [1] admits no closed timelike  geodesics, even though the spacetime includes
closed timelike curves. In the case of the  Kerr metric,  Carter [3] proved that closed timelike
paths exist for some parameter values of the  metric.  Calvani et al.  [7] then showed that a class of
timelike geodesics in the G\"odel spacetime  do not violate causality, but  de Felice and Calvani [8]
later found numerical solutions for null geodesics that violate causality under  ``very particular
conditions.''  More recently, Steadman [9] found closed, circular  timelike geodesics in the  van
Stockum spacetime [10] for a rotating infinite dust cylinder, along which coordinate time runs backward.

In this paper, we consider Bonnor's axially symmetric rotating dust cloud metric [11], on which a closed 
(but nongeodesic) null path has already been observed to exist [12]. We analyze the causality violating
region and give numerical  plots of timelike geodesics along which coordinate time runs
backward on portions of the  paths.  We show that some of the geodesics connect the causality violating
region to asymptotically flat spatial infinity.
   
In Section 2, we introduce notation and describe the Bonnor dust cloud metric. In Section 3,
we prove that the causality violating region for this spacetime includes the interior of a torus centered
at the origin of coordinates, and we use that information to 
produce numerical plots of causality violating geodesics.  Concluding remarks are given
Section 4. Finally in the Appendix, we prove that no 
timelike or null circular closed
geodesics, centered  on the axis of symmetry of the spacetime (analogous to geodesics considered by 
Steadman [9]), violate
causality.\\

\noindent\textbf{{\normalsize 2. BONNOR'S DUST METRIC}}\\

\indent A solution to the field equations given by Bonnor [11] is an axially symmetric metric
which  describes a cloud of rigidly rotating dust particles moving along circular geodesics about the 
z-axis in hypersurfaces of $z=\mbox{constant}$. The line element is
\begin{equation}
ds^{2}=-dt^{2}+(r^{2}-n^{2})d\phi^{2}+2ndtd\phi+e^{\mu}(dr^{2}+dz^{2})\;,
\end{equation}
where, in Bonnor's comoving (i.e., corotating) coordinates,
\begin{equation}
n=\frac{2hr^{2}}{R^{3}}\,,\;\;\;\;\;\mu=\frac{h^{2}r^{2}(r^{2}-8z^{2})}{2R^{8}}\,,\;\;\;\;\;R^{2}=
r^{2}+z^{2}\,,
\end{equation}
$h$ is a rotation parameter, and we have the gauge condition
\begin{equation}
(g_{t\phi})^{2}-g_{tt}g_{\phi\phi}=r^{2}\,.
\end{equation}
Bonnor's metric has an isolated singularity at $r=z=0$.
\noindent The energy density $\rho$ is given by
\begin{equation}
8\pi \rho=\frac{4e^{-\mu}h^{2}(r^{2}+4z^{2})}{R^{8}}\,.\\
\end{equation}
\noindent As $R\rightarrow \infty$, $\rho$ approaches zero rapidly and the metric 
coefficients tend to Minkowski values. Moreover, all the Riemann
curvature tensor elements vanish at spatial infinity.\\

\noindent \textbf{{\normalsize 3. CAUSALITY VIOLATING TIMELIKE GEODESICS}}\\

\indent From Eqs. (1) and (2) we obtain the Hamiltonian
\begin{eqnarray}
H&=&\frac{1}{2}g^{\alpha\beta}p_{\alpha}p_{\beta}\nonumber\,,\\
&=&\frac{(n^{2}-r^{2})}{2r^{2}}p_{t}^{2}+\frac{1}{2r^{2}}p_{\phi}^{2}+\frac{n}{r^{2}}p_{t}p_{\phi}+
\frac{1}{2}e^{-\mu}(p_{r}^{2}+p_{z}^{2})\,,
\end{eqnarray}
where the angular momentum $p_{\phi}$ and the energy $E=-p_{t}$ are conserved quantities.
Since for timelike geodesics $H=-1/2$, it follows from Eq. (5) that timelike geodesics with angular
momentum $p_{\phi}$ and energy $E$ can exist only in the region $S_{B}$ of spacetime given by
\begin{equation}
S_{B}=\{(t,\phi, r,z)|-p^{2}_{\phi}+2nEp_{\phi}+(r^{2}-n^{2})E^{2}=
r^{2}(1+e^{-\mu}(p_{r}^{2}+p_{z}^{2}))\}\,.
\end{equation}\\
\noindent \textbf{Remark:} Steadman [13] investigated the ``allowed region'' for null geodesics in
Bonnor's metric.  He showed that the region corresponding to $S_{B}$, for null geodesics, is
topologically connected for $p_{\phi}/E \leq 2\sqrt{2h}$ and has two components otherwise. 
So for $p_{\phi}/E > 2\sqrt{2h}$, null geodesics may be trapped in Steadman's ``central region of
confinement,'' i.e., in the component with smaller $r$ values.  We find that the situation is
qualitatively the same for timelike geodesics, except that in this case the topology of the physically
allowed region depends on both $p_{\phi}$ and $E$ and not just their ratio.\\

To find causality violating timelike geodesics, we consider Hamilton's equation for 
$\dot{t}$ for an arbitrary geodesic, where the overdot represents differentiation with respect
to proper time, and for convenience, we set $h=1$. 
\begin{equation}
\dot{t}=E+\frac{2p_{\phi}}{(r^{2}+z^{2})^{3/2}}-\frac{4Er^{2}}{(r^{2}+z^{2})^{3}}
\end{equation}

\noindent The following proposition shows that a causality violating region exists for 
this spacetime.\\

\noindent \textbf{Proposition 1}:  \textit{There exists a spacelike torus $T$ 
centered at $r=z=0$ in the Bonnor spacetime such that $\dot{t}<0$ along any timelike geodesic in the
interior of $T$.} \\

\noindent\textbf{Proof}. In the plane $z=0$, Eq. (7) becomes,
\begin{equation}
\dot{t}=\frac{2p_{\phi}}{r^{3}}-\frac{E(4-r^{4})}{r^{4}}\,.
\end{equation}
For $z=0$ and $\dot{t}<0$, it follows that we must have
\begin{equation}
-\left[(r^{4}-4)E+2 r p_{\phi}\right]>0\,.
\end{equation}
Now substituting $z=0\,,p_{z}=0\,,h=1$ and Eqs. (2) into the equation in (6) for the physically
allowed  region for timelike geodesics gives,
\begin{equation}
p_{r}^{2}
=\frac{1}{r^{4}}\exp\left(\frac{1}{2r^{4}}\right)[(E^{2}-1)r^{4}-(2E-rp_{\phi})^{2}].
\end{equation}
Solutions to (10) exist provided that,
\begin{equation}
(E^{2}-1)r^{4}-(2E-rp_{\phi})^{2}\geq 0.
\end{equation}
It follows that for $z=0$, timelike geodesics with $\dot{t}<0$ exist exactly for those parameters
and coordinates for which the inequalities (9) and (11) are simultaneously satisfied. It is easy to check
that the solution set is nonempty with strict inequality in (11) (see the numerical examples below). 
From standard continuity arguments, it follows that $\dot{t}<0$ for all timelike  geodesics with
parameters and coordinates  in an open set in $(r, \phi, z, E, p_{\phi})$ space. Since (6) and (7) are
independent of $\phi$ this open set is independent of $\phi$.  This completes the proof.\\

\noindent \textbf{Remark:}  From inequality (11) it follows that we must have $E\geq 1$.  It is easy to 
see that inequalities (9) and (11) cannot be satisfied for $p_{\phi}=0$ since inequality (9) then
requires $r<\sqrt{2}$, while inequality (11) requires $r\geq \sqrt{2}$. It is also easy to show that the 
inequalities cannot be satisfied for $p_{\phi}=-|p_{\phi}|<0$. Multiplying inequality (9) by 
$E$ yields $A\equiv -E^{2}r^{4}+4E^{2}+2E|p_{\phi}|r>0$.  Inequality (11) may now be written
as $-A-r^{4}-2E|p_{\phi}|r-|p_{\phi}|^{2}r^{2}\geq0$ and it is clearly incompatible with $A>0$. From 
inequality (9) we see that for $p_{\phi}>0$, we must have $r<\sqrt{2}$.  Thus in the
plane $z=0$, we must have $r<\sqrt{2}$, $p_{\phi}>0$, and $E\geq 1$ for any portion of a timelike
geodesic along which $\dot t<0$.\\

\indent  Using the Proposition and the Remark immediately above, we now give two
numerical examples of timelike causality violating geodesics (CVG), both lying in the $z=0$ plane with
$p_{z}=0$. One is a spatially bound, evidently quasiperiodic geodesic, and the other is an unbound
geodesic.  In the quasiperiodic example (Figures 1, 2, and 3) we let $E=1.3,\;p_{\phi}=3$, then
inequalities (9) and (11) are both satisfied for $r\in[0.722235...,0.784571...)$.  The initial values
for the timelike CVG of Figure 1 are
$r=1$, $p_{r}=0.934785...$.  For the unbound CVG example (Figures 4, 5, and 6) we have $E=4.12078...$,
$p_{\phi}=2.5$, then inequalities (9) and (11) are both satisfied for $ r\in[1.15680...,1.25462...)$. 
The initial values for the timelike CVG of Figure 4 are $r=4$, $p_{r}=-4$.\\

\noindent \textbf{{\normalsize 4. CONCLUDING REMARKS}}\\

Bonnor's dust metric includes a causality violating region with nonempty interior. Coordinate time 
runs backward along
the portions of all timelike geodesics within this region. 
Bonnor [14] has argued that closed timelike paths are possible, not only in
physically unrealistic solutions to the Einstein field equations, but also for solutions that ``refer to 
ordinary materials in situations which might occur in the laboratory, or 
in astrophysics.''  He has called for new interpretations of this phenomenon. The spacetime considered
here has 
some unrealistic features. It has an isolated singularity with no event horizon. However, the 
singularity is inaccessible to all null geodesics except possibly along the axis of symmetry of the
spacetime, a set of zero measure. In addition, the metric 
coefficients for this spacetime tend to Minkowski values at spatial infinity and all the Riemann
curvature tensor elements vanish there. Numerical calculations show that the causality violating region 
of this spacetime is accessible to timelike geodesic paths from distant, 
asymptotically flat regions of the spacetime. Thus the causality violating region may be reached by
an observer with no expenditure of energy. A new physical interpretation of closed timelike curves may 
have interesting implications for this spacetime. \\

\noindent \textbf{{\normalsize APPENDIX}}\\

In this Appendix we show that Bonnor's dust metric does not have timelike or null circular 
geodesics about the z-axis in hypersurfaces of $z=\mbox{constant}$ with $\dot{t}\leq 0$.  We set 
the parameter $h=1$ and shall assume that $E>0$. Hamilton's equation for $\dot{t}$ is Eq. (7). 
Hamilton's equation for $\dot{\phi}$ is:

\begin{equation}
\dot{\phi}=\frac{p_{\phi}}{r^{2}}-\frac{2E}{(r^{2}+z^{2})^{3/2}}
\end{equation}\\
If we let $p_{r}=p_{z}=0$ in Hamilton's equations for $\dot{p}_{r}$ and $\dot{p}_{z}$, we have
\begin{eqnarray}
\dot{p}_{r}&=&\frac{p_{\phi}^{2}}{r^{3}}-\frac{6Ep_{\phi}r}{(r^{2}+z^{2})^{5/2}}-\frac{E^{2}
(4-3(r^{2}+z^{2})^{2})r}{(r^{2}+z^{2})^{3}}\nonumber\\
           & &+\frac{3E^{2}(4r^{2}-(r^{2}+z^{2})^{3})r}{(r^{2}+z^{2})^{4}}\,,\\
\dot{p}_{z}&=&-\frac{6Ep_{\phi}z}{(r^{2}+z^{2})^{5/2}}+\frac{3E^{2}z}{r^{2}+z^{2}}
+\frac{3E^{2}(4r^{2}-(r^{2}+z^{2})^{3})z}{(r^{2}+z^{2})^{4}}\,.
\end{eqnarray}\\
Equating the above equations to zero and solving for $r$ and $z$ we obtain the following
solutions for $z$ real and $r>0$:
\begin{eqnarray}
z&=&0\,,\;\;\;\;r=\frac{2E}{p_{\phi}}\,,\\
z&=&0\,,\;\;\;\;r=\frac{4E}{p_{\phi}}\,,\\
z&=&\pm\left(\left(\frac{2Er^{2}}{p_{\phi}}\right)^{\frac{2}{3}}-r^{2}\right)^{\frac{1}{2}}\,.
\end{eqnarray}\\
Solution (17) reduces to solution (15) if we require that $z=0$. The above solutions will be timelike
circular geodesics provided $p_{r}=p_{z}=0$.  To check this we introduce the function $K$
below,
\begin{equation}
e^{\mu}(H+\frac{1}{2})=\frac{1}{2}(p_{r}^{2}+p_{z}^{2})+\frac{e^{\mu}}{2}K=0\,,
\end{equation}
where $H$ is the Hamiltonian of Eq. (5), and using Eqs. (2), we have
\begin{equation}
K(r,z,E,p_{\phi})=1+\frac{p_{\phi}^{2}}{r^{2}}-\frac{4Ep_{\phi}}{(r^{2}+z^{2})^{3/2}}+
\left(\frac{4r^{2}}{(r^{2}+z^{2})^{3}}-1\right)E^{2}\,.
\end{equation}
From Eq. (18) we see that for circular timelike geodesics $K$ must vanish so that $p_{r}=p_{z}=0$.
Substituting solution (15) ($z=0$) or solution (17) ($z\neq0$) in $K$ we find that $K=0$
forces $E=1$. Furthermore if we substitute  solution (15) or (17) into Eq. (12) for
$\dot{\phi}$ and Eq. (7) for $\dot{t}$, we get $\dot{\phi}=0$ and $\dot{t}=E$,  therefore the circular
geodesics of solutions (15) and (17) correspond to the paths of Bonnor's dust particles.  If we
substitute solution (16) in $K$ we find that $E=((1+\sqrt{1+(p_{\phi}/2)^{4}})/2)^{1/2}$ and this time
Eqs. (12) and (7) give
\begin{eqnarray}
\dot{\phi}&=&\frac{p_{\phi}^{3}}{32E^{2}}\,,\\
\dot{t}&=&E+\frac{p_{\phi}^{4}}{64E^{2}}\,.
\end{eqnarray}\\
We see from Eq. (21) that $\dot{t}>0$ for these geodesics also.

Finally, repeating the above calculations for null geodesics we find that only solution (16) can be
a null circular geodesic provided that $E=p_{\phi}/(2\sqrt{2})$ and in this case also $\dot{t}>0$, 
where the overdot now represents differentiation with respect to an affine parameter.\\

\noindent \textbf{ACKNOWLEDGMENTS}\\

\noindent The authors wish to thank Professors Cristina Cadavid and John Lawrence for helpful comments.\\

\noindent \textbf{{\normalsize REFERENCES}}

\begin{enumerate}
\def\labelenumi{[\theenumi]}
\item G\"odel, K. (1949). \textit {Rev. Modern Phys.}, \textbf {21}, 447; reprinted in (2000) 
Gen. Rel. Grav. \textbf{32}, 7409.
\item Tipler, F. J. (1974). \textit{Phys. Rev. D} \textbf{9}, 2203.
\item Carter, B. (1968). \textit {Phys. Rev.} \textbf{174}, 1559.
\item Bonnor, W. B. (2002). \textit{Class. Quantum Grav.}  \textbf{19}, 5951.
\item Visser, M. (1996).  \textit{Lorentzian Wormholes}, Springer-Verlag, New York.
\item Chandrasekhar, S. and Wright, L. (1961). \textit{Proc. Natl. Acad. Sci. USA,} \textbf{48}, 341.
\item Calvani, M., de Felice, F., Muchotrzeb, B., and Salmistraro, F. (1978). \textit{Gen Rel. Grav.} 
\textbf{9}, 155.
\item de Felice, Calvani, M. (1978).  \textit{Gen Rel. Grav.} \textbf{10}, 335.
\item Steadman, B. R. (2003). \textit{Gen. Rel. Grav.} \textbf{35}, 1721.
\item van Stockum, W. J. (1937). \textit{Proc. R. Soc. Edin.} \textbf{57}, 135.
\item Bonnor, W. B. (1977). \textit{J. Phys. A: Math. Gen.} \textbf{10}, 1673.
\item Collas, P. and Klein, D. (2004). \textit{Gen. Rel. Grav. }\textbf{36}, 1197. 
\item Steadman, B. R. (1999). \textit{Class. Quantum Grav.} \textbf{16}, 3685.
\item Bonnor, W. B. (2003). \textit{Int. J. Mod. Phys. D} \textbf{12}, 1705.
\end{enumerate}

\end{document}